\documentclass[11pt]{article}
\usepackage{amsmath}
\usepackage{amsfonts}
\usepackage{amssymb}
\usepackage{hyperref}
\usepackage{xcolor}
\usepackage[margin=2.5cm]{geometry}
\usepackage{extarrows,mathtools,graphicx,subfigure,setspace,bbold}
\usepackage{cite}
\usepackage{slashed}
\usepackage{tikz}
\usepackage{arydshln,leftidx}
\usepackage{authblk}

\numberwithin{equation}{section}
\newcommand{\be}{\begin{equation}}
	\newcommand{\ee}{\end{equation}}
\newcommand{\ba}{\begin{align}}
	\newcommand{\ea}{\end{align}}

\hypersetup{colorlinks=true, linkcolor=blue, citecolor=magenta, linktoc=page}

\begin{document}
	\onehalfspacing
\begin{titlepage}
	\thispagestyle{empty}

	\vspace{.4cm}
\begin{center}{\Large \textbf{
Holographic Krylov Complexity with Lifshitz Scaling and Hyperscaling Violation
}}\end{center}

\vspace*{15mm}
		\vspace*{1mm}

\begin{center}
{Kazem Bitaghsir Fadafan$^a$ and  M. Reza Mohammadi Mozaffar$^b$ 
}
\end{center}
\vspace*{1cm}

\begin{center}
{\it $^a$ Faculty of Physics, Shahrood University of Technology,  P.O.Box 3619995161, Shahrood, Iran\\
$^b$ Department of Physics, University of Guilan, P.O. Box 41335-1914, Rasht, Iran\\
		}
		
		\vspace*{0.5cm}
		{E-mails: {\tt bitaghsir@shahroodut.ac.ir, mmohammadi@guilan.ac.ir}}%
\end{center}

\begin{abstract}
Following the holographic proposal that identifies the growth rate of Krylov complexity with the proper radial momentum of an infalling massive probe, we study Krylov complexity in Lifshitz and hyperscaling-violating backgrounds. For pure Lifshitz geometries, we find quadratic complexity growth for all values of the dynamical exponent. For hyperscaling-violating backgrounds, we show that the hyperscaling-violating exponent directly controls the late-time growth exponent, with a special limiting case exhibiting oscillatory behavior with a logarithmic envelope. We also extend our analysis to top-down string constructions, confirming the robustness of the scaling behaviors obtained in the bottom-up approach. Our analysis establishes that the momentum-Krylov correspondence extends naturally to non-relativistic holographic settings and remains well-defined despite the causal pathologies of Lifshitz spacetimes.
\end{abstract}

\end{titlepage}
\newpage

\tableofcontents
\noindent
\hrulefill

%%%%%%%%%%%%%%%%%%%%%%%%%%
	
\section{Introduction}

The gauge/gravity duality, originating from the AdS/CFT correspondence~\cite{Maldacena:1997re,Gubser:1998bc,Witten:1998qj}, establishes a precise geometric dictionary for strongly coupled quantum field theories, allowing non-perturbative phenomena to be studied using classical gravitational methods in one higher dimension. Within this framework, the holographic computation of quantum information-theoretic quantities has become an active research direction. Measures such as entanglement entropy, mutual information, out-of-time-order correlators (OTOCs), and quantum complexity have been given well-defined gravitational duals~\cite{Ryu:2006bv,Hubeny:2007xt,VanRaamsdonk:2010pw,Maldacena:2016upp,Susskind:2014rva,Susskind:2014moa,Brown:2015lvg}. These correspondences have provided insights into the emergence of spacetime geometry from entanglement, as well as into black hole interiors, quantum chaos, and thermalization dynamics.

Among these, quantum complexity has received considerable attention in holography. Two prominent proposals have emerged: the complexity-volume (CV) duality, which identifies the complexity of the boundary state with the volume of a maximal codimension-one bulk surface ending on the boundary~\cite{Susskind:2014rva,Susskind:2014moa}, and the complexity-action (CA) duality, which equates complexity to the gravitational action evaluated on a bulk region known as the Wheeler-DeWitt patch~\cite{Brown:2015lvg}. Both proposals have successfully reproduced expected features of complexity, such as linear growth at late times for black hole backgrounds, and have been extensively studied in various holographic settings. However, they suffer from certain ambiguities, including dependence on the choice of reference state and, in the CV case, on the specific prescription for the codimension-one surface.

A more recent development in the study of quantum many-body dynamics is Krylov complexity, a measure of operator growth~\cite{Parker:2018yvk}. Krylov complexity quantifies the spread of an initially simple operator under time evolution within an orthonormal Krylov basis, constructed recursively via the Lanczos algorithm. The complexity, defined as the expectation value of the Krylov position operator, serves as a diagnostic of integrability versus chaos, operator scrambling, and thermalization in systems ranging from quantum spin chains and random matrix theory to quantum field theories~\cite{Bhattacharjee:2022vlt,Bhattacharya:2022gbz,Rabinovici2022,Bhattacharjee:2022qjw,Bhattacharjee:2022lzy,Alishahiha:2022anw,Avdoshkin2022,Camargo:2022,Bhattacharyya:2023grv,Adhikari:2023evu,Bhattacharjee:2023uwx,Alishahiha:2024rwm,Nandy:2024htc,Caputa:2026hvh}. A precise holographic dual for Krylov complexity has been proposed, linking boundary operator dynamics to bulk geodesic motion~\cite{Caputa:2024sux}. In this proposal, the time derivative of the Krylov complexity $C_K(t)$ in the boundary theory is identified with the proper radial momentum of a massive probe particle moving along a geodesic in the dual bulk geometry
\begin{equation}\label{CK}
\dot{C}_K(t) = -P_\rho,
\end{equation}
where $\rho$ is the proper radial coordinate. This geometric interpretation has been tested in several relativistic holographic backgrounds, including AdS black holes, confining geometries, and the Coulomb branch of $\mathcal{N}=4$ super Yang-Mills theory \cite{Caputa:2024sux,Das:2024tnw,Fan:2024iop,Fatemiabhari:2025usn,Fatemiabhari:2026goj,Zoakos:2026obl,Nastase:2026lhz}. These studies have revealed universal features of complexity growth, such as early-time linear or exponential behavior and late-time saturation, often connected to black hole horizons and chaos bounds. The Krylov complexity prescription differs from CV and CA in both concept and scope. CV and CA are holographic conjectures that assign a complexity to the boundary quantum state via bulk geometric quantities. Krylov complexity, by contrast, is a field-theoretic quantity defined from operator growth dynamics, with its holographic dual given by the proper radial momentum of an infalling probe. This provides a state-independent, ambiguity-free geometric dual that is intrinsically dynamical.

Given these advantages, it is natural to ask whether the holographic Krylov complexity prescription applies to non-relativistic bulk backgrounds, whose dual field theories lack Lorentz invariance. This is the central question we address in this work. Specifically, we investigate the scaling and time evolution of Krylov complexity in theories where Lorentz invariance is broken. Such systems are physically well-motivated: many strongly coupled systems in condensed matter physics, including quantum critical points in correlated materials, ultracold atomic gases, and certain non-relativistic gauge theories, exhibit anisotropic scaling symmetries. Their holographic descriptions are provided by Lifshitz geometries, often with hyperscaling violation~\cite{Kachru:2008yh,Taylor:2015tka,Kiritsis:2015oxa}, which are invariant under the generalized scaling transformation
\begin{equation}\label{scaling}
t \to \lambda^z t, \qquad x^i \to \lambda x^i, \qquad r \to \lambda^{-1} r, \qquad ds \to \lambda^{\theta/d} ds,
\end{equation}
where $z$ is the dynamical critical exponent, $\theta$ is the hyperscaling-violating exponent, and $d$ is the spatial dimension. The effective dimensionality of the system is given by $d_{\text{eff}} = d - \theta$. These geometries arise naturally in Einstein-dilaton-Maxwell theories and string theory embeddings, providing a controlled setting for non-relativistic holography~\cite{Hartnoll:2016apf}.

Extending the holographic formulation of Krylov complexity to these non-relativistic backgrounds is timely. 
In the conformal case, Ref.~\cite{Caputa:2024sux} established a sharp relation between the growth rate of spread complexity in two-dimensional CFTs and the proper radial momentum of particles in AdS$_3$, relying crucially on the Virasoro/SL(2,$\mathbb{R}$) symmetry of the dual field theory. However, this approach is not applicable to Lifshitz or hyperscaling-violating backgrounds, where conformal invariance is broken. While field-theory results exist for models with nontrivial $z$ (with $\theta = 0$) in the literature, see e.g. Refs.~\cite{Vasli:2023,Imani:2025etp}, the methods employed there are distinct and do not generalize to the full $(z,\theta)$ parameter space. Our work fills this gap by providing a holographic computation of Krylov complexity in such non-relativistic backgrounds, thereby extending the scope of the momentum-Krylov correspondence beyond the conformal regime. It is worth noting that the analyses of Refs.~\cite{Vasli:2023,Imani:2025etp} are performed in free Lifshitz field theories, which are fundamentally different from a holographic dual, which typically describes a strongly coupled field theory. Therefore, a direct comparison between our holographic results and those free-field computations is not straightforward, as they probe different regimes of the theory space. Nevertheless, to further support our holographic findings, we have performed a complementary field-theory analysis of Krylov complexity in Lifshitz scalar and Dirac theories at early times, finding the same universal quadratic growth, consistent with our holographic results.

Before proceeding, it is worth clarifying the distinction between our work and certain existing holographic studies of complexity in Lifshitz and hyperscaling-violating backgrounds. In particular, Ref.~\cite{Alishahiha:2018tep} investigated computational (circuit) complexity in Lifshitz and hyperscaling-violating geometries using the ``complexity=action'' proposal. While that work shares the same class of gravitational backgrounds, it addresses a fundamentally different notion of complexity: circuit complexity measures the minimal number of quantum gates required to prepare a given state, whereas Krylov complexity quantifies the spread of an operator in the Krylov basis and is intrinsically tied to operator growth dynamics. These two quantities probe different aspects of the dual field theory and are not directly comparable. Our work focuses exclusively on Krylov complexity within the momentum-Krylov correspondence, and we do not attempt to compare or connect our results to circuit complexity computations. We simply note the existence of these complementary holographic studies for completeness.

The dynamical exponent $z$ controls the relative scaling of time and space, affecting dispersion relations, transport, and information scrambling. The hyperscaling-violating exponent $\theta$ introduces additional scaling dimensions that can alter critical behavior. Understanding how these parameters influence operator growth may reveal new universal features of quantum chaos and complexity in anisotropic systems and possibly lead to generalized bounds on complexity growth rates beyond the relativistic regime \cite{Maldacena:2015waa}. In this direction, quantum chaos, entanglement measures, and computational complexity in Lifshitz theories have been extensively studied in recent years on the field theory side, leading to a rich variety of new results \cite{Ogawa2012,MohammadiMozaffar:2017nri, He:2017wla, Gentle:2017ywk, MohammadiMozaffar:2017chk, MohammadiMozaffar:2018vmk, MohammadiMozaffar:2019gpn, Hartmann:2021vrt, Mozaffar:2021nex, Mintchev:2022xqh, Mintchev:2022yuo,Vasli:2023,Imani:2025etp,MohammadiMozaffar:2026osc}.

Holographic Krylov (spread) complexity has been extensively tested across a variety of backgrounds~\cite{Fatemiabhari:2025usn,Fatemiabhari:2026goj,Zoakos:2026obl,Nastase:2026lhz}. In confining holographic geometries with a smooth infrared end-of-space, a robust and universal feature emerges: Krylov complexity exhibits oscillatory behaviour, with frequency controlled by the confinement scale. 
 These oscillations have been shown to qualitatively match field-theoretic results from the longitudinally perturbed Ising model, suggesting they constitute a universal signature of confinement and a sensitive probe of infrared reorganisation in strongly coupled quantum field theories~\cite{Fatemiabhari:2025usn,Fatemiabhari:2026goj}. Recent studies have extended this framework to higher-dimensional superconformal field theories and top-down Type~IIB solutions describing twisted-circle compactifications of $\mathcal{N}=4$ SYM~\cite{Zoakos:2026obl,Nastase:2026lhz,Fatemiabhari:2026rob}. For example, in strongly coupled six-dimensional superconformal field theories with holographic duals in massive Type~IIA supergravity~\cite{Fatemiabhari:2026rob}, the dynamics of massive geodesic probes have been analysed, allowing the bulk particle to move along radial and internal directions. In the dual field theory, these motions encode operator growth and the associated global symmetries.
 
In this work, we systematically develop the holographic computation of Krylov complexity in geometries with Lifshitz scaling and hyperscaling violation. It is worth noting that Lifshitz and Schr\"odinger solutions in higher-dimensional supergravity typically do not accommodate simple geodesic motion along internal directions. We have verified, for instance, that in the supergravity solutions of Ref.~\cite{Donos:2010tu}, extending the analysis to include motion along the internal space is not straightforward in the context of purely radial motion. We therefore restrict our attention to geometries with Lifshitz scaling and hyperscaling violation, where we consider massive probe particles and derive the conserved quantities associated with the Killing symmetries of the background. Solving the geodesic equations, we obtain explicit expressions for the canonical and proper radial momenta, from which we compute the time dependence of $\dot{C}_K(t)$ and analyze its dependence on $z$, $\theta$, and the probe mass $m$.

To strengthen the holographic interpretation and address the limitations of the bottom-up approach, we will also consider explicit top-down constructions derived from string. We will review two classes of such solutions: Lifshitz-like fixed points in type IIB supergravity arising from D3-D7 systems, and intersecting F-Dp brane configurations that yield Lifshitz spacetimes with hyperscaling violation. For both classes, we will derive the geodesic motion of a massive probe and compute the corresponding Krylov complexity growth, and compare the results with the bottom-up analysis. This comparison will confirm that the scaling behaviour of Krylov complexity is a robust feature of non-relativistic holographic backgrounds, independent of the specific UV completion. The top-down constructions will also provide a microscopic interpretation and open the door to future studies of extended probes and symmetry-resolved complexity.

The paper is organized as follows. In Section~\ref{generalframe}, we review the holographic proposal for Krylov complexity and establish the general framework for computing proper radial momentum in arbitrary bulk backgrounds. We then introduce the Lifshitz and hyperscaling-violating geometries that will be the focus of our analysis, and discuss the constraints imposed by the null energy condition. Section~\ref{lifback} is devoted to pure Lifshitz backgrounds: we derive the geodesic equations for a massive infalling probe, compute the canonical and proper radial momenta, and obtain the time evolution of Krylov complexity. We also discuss the pathologies of Lifshitz spacetimes and explain why they do not affect our calculation. In addition, we compare our holographic results with a field-theory analysis of Krylov complexity in Lifshitz scalar and Dirac theories, confirming the universal quadratic growth at early times. Section~\ref{hvback} generalizes the analysis to hyperscaling-violating geometries, where we derive the trajectory and momenta in terms of hypergeometric functions, and provide systematic early-time and late-time asymptotic expansions. Section~\ref{topdown} presents the top-down constructions, where we review Lifshitz fixed points from string theory and hyperscaling-violating solutions from intersecting branes, and compute Krylov complexity in these backgrounds. Section~\ref{conclusion} summarizes our main findings and outlines promising future directions, including extensions to probes with internal structure and connections to other quantum information measures in non-relativistic holography.

\section{General framework}\label{generalframe}
In the following sections, we apply the holographic prescription of \cite{Caputa:2024sux} to determine the time evolution of Krylov complexity in the boundary theory using Eq.~\eqref{CK}. To establish the general framework, consider a bulk background with a metric of the form
\begin{equation}
ds^2 = G_{00}(r) dt^2 + G_{rr}(r) dr^2 + G_{ij}(r) dx^i dx^j + \text{(internal space)}, \label{eq:general_metric}
\end{equation}
where \(r\) is the radial holographic coordinate, with the conformal boundary located at \(r \to 0\). We assume the metric components depend only on \(r\), reflecting translational invariance in the boundary spatial directions \(x^i\) and time \(t\). We discuss the effect of the internal space at the end of this section.

To compute Krylov complexity, we introduce a proper distance coordinate \(\rho\) along the radial direction. The proper radial distance is defined by
\begin{equation}\label{eq:proper_coordinate}
d\rho^2 = G_{rr}(r) dr^2. 
\end{equation}
This coordinate change is essential because the holographic proposal identifies the growth rate of Krylov complexity with the proper radial momentum rather than the coordinate momentum \cite{Caputa:2024sux}. We consider an infalling probe --- which may be a particle, string, or brane --- moving along a radial geodesic \cite{Nastase:2026lhz}. The probe is subject to the initial conditions
\begin{equation}\label{inicon}
r(0) = \epsilon, \qquad \dot{r}(0) = 0,
\end{equation}
where \(\epsilon\) is a UV cutoff near the boundary. These conditions correspond to releasing the probe from rest at the UV boundary at time \(t=0\). The probe then falls inward under the influence of the bulk gravitational field.

Let \(\mathcal{L}\) denote the Lagrangian of the probe. The canonical momentum conjugate to the proper radial coordinate \(\rho\) is given by
\begin{equation}
P_\rho = \frac{\partial \mathcal{L}}{\partial \dot{\rho}}, \label{eq:canonical_momentum}
\end{equation}
where the dot denotes differentiation with respect to the affine parameter (typically proper time). This momentum is conserved if the background possesses appropriate Killing symmetries; in particular, time translation invariance yields conservation of the energy \(E = - \partial \mathcal{L} / \partial \dot{t}\).

According to the holographic proposal for Krylov complexity, the rate of change of Krylov complexity in the boundary theory is identified with the proper radial momentum of the infalling probe~\cite{Caputa:2024sux}
\begin{equation}
\dot{C}_K(t) = - P_\rho(t). \label{eq:complexity_rate}
\end{equation}
The negative sign ensures that complexity grows as the probe falls inward (since \(P_\rho\) is negative for infalling motion). Thus, by solving the geodesic equations for the probe and computing \(P_\rho\) as a function of boundary time \(t\), we obtain the time evolution of Krylov complexity in the dual field theory.

We now specialize to backgrounds with a general dynamical critical exponent $z$ and hyperscaling violation exponent $\theta$. These theories admit a fixed point invariant under the anisotropic scaling transformation in Eq.~\eqref{scaling}. A number of holographic aspects of these theories have been studied in \cite{Dong:2012se,Alishahiha:2012qu,BitaghsirFadafan:2015asm,Gath:2012pg,Alishahiha:2014cwa}. Consider a $(d+2)$-dimensional solution in Einstein gravity coupled to appropriate matter fields (see, e.g., \cite{Dong:2012se})
\begin{equation}\label{eq:metricNR}
ds^2=\frac{1}{r^{2(d-\theta)/d}} \left( -\frac{1}{r^{2(z-1)}} dt^2 + dr^2 + d\vec{x}^2 \right).
\end{equation}
As noted in \cite{Dong:2012se}, the null energy condition imposes the bounds
\begin{equation}
(d-\theta)(d(z-1)-\theta) \ge 0, \qquad (z-1)(d-\theta+z) \ge 0, \label{eq:NEC}
\end{equation}
also, entanglement entropy analysis and string theory realizations of hyperscaling-violating geometries suggest inconsistency when $\theta > d$. Further, the case $\theta = d-1$ has been identified as a particularly interesting holographic model, offering a gravitational dual for a theory with a Fermi surface, as evidenced by its leading large-$N$ thermodynamic scaling \cite{Ogawa2012}. 

Note that the metric in Eq.~\eqref{eq:general_metric} generically includes an internal space. The role of such internal spaces in holographic Krylov complexity was investigated in Ref.~\cite{Nastase:2026lhz} for a massive particle in AdS$_5 \times S^5$ carrying conserved R-charge. It was shown that motion in the internal space modifies complexity growth, providing a holographic realization of symmetry-resolved Krylov complexity. Since our background in Eq.~\eqref{eq:metricNR} does not include an internal space, we do not consider such effects in this work.

Now equipped with all the necessary tools, we proceed to calculate the time evolution of Krylov complexity. Unfortunately, it is not possible to obtain an analytic expression for the growth rate in a general setup. In the following, we present numerical results along with analytic treatments for specific cases.

\section{Lifshitz background}\label{lifback}

We now specialize to a Lifshitz geometry with a nontrivial dynamical exponent $z$, and study Krylov complexity using the holographic prescription outlined above. We first analyze the geodesic motion of an infalling massive particle, and then compute the associated proper momentum, from which the complexity growth rate follows via Eq.~\eqref{eq:complexity_rate}. 

\subsection{Geodesic motion of an infalling particle}

We consider a massive particle moving in a Lifshitz spacetime, which is dual to a boundary state with nonrelativistic scaling symmetry. The metric is obtained by setting $\theta = 0$ in Eq.~\eqref{eq:metricNR}
\begin{equation}
ds^2=-\frac{dt^2}{r^{2z}}+\frac{dr^2+d\vec{x}^2}{r^2}. \label{eq:lifshitz_metric}
\end{equation}
For radial geodesics, we set $d\vec{x}=0$. Defining the proper time interval as $d\tau^2=-ds^2$ and using the above equation, we obtain
\begin{equation}
d\tau=\frac{dt}{r^z}\sqrt{1-r^{2z-2}\dot{r}^2}, \label{eq:dtau}
\end{equation}
where the dot denotes differentiation with respect to $t$. Note that the proper time is real only when the expression under the square root is non-negative, i.e., $r^{2z-2} \dot{r}^2 \leq 1$, which imposes a constraint on the allowed radial velocities. In the relativistic limit $z=1$, this condition reduces to the familiar bound $\dot{r}^2\leq 1$. Using the relation above, the action for a massive particle of mass $m$ takes the form
\begin{equation}\label{eq:action_lifshitz}
S=-m \int \frac{dt}{r^z} \sqrt{1 - r^{2z-2} \dot{r}^2} \equiv \int \mathcal{L}(r, \dot{r}) \, dt. 
\end{equation}
Since the Lagrangian has no explicit time dependence, the corresponding Hamiltonian (energy) is conserved
\begin{equation}
H=\dot{r} \frac{\partial \mathcal{L}}{\partial \dot{r}}-\mathcal{L}=\text{constant}. \label{eq:hamiltonian_conserved}
\end{equation}
Denoting this constant energy by $E$, a straightforward calculation yields
\begin{equation}
E=\frac{m}{r^z \sqrt{1-r^{2z-2}\dot{r}^2}}=\frac{m}{\epsilon^z}, \label{eq:E_lifshitz}
\end{equation}
where in the second equality we have used the initial conditions in Eq.~\eqref{inicon} to express the energy in terms of the UV cutoff $\epsilon$. Solving Eq.~\eqref{eq:E_lifshitz} for $\dot{r}^2$, we obtain the radial velocity squared
\begin{equation}
\dot{r}^2=r^{2-2z} \left(1-\left(\frac{\epsilon}{r}\right)^{2z}\right). \label{eq:zdot2_lifshitz}
\end{equation}
We always choose the positive square root, corresponding to the particle falling inward from the UV boundary ($\dot{r}>0$ for $t>0$). For the relativistic case $z=1$, this reduces to the familiar AdS result $\dot{r}^2=1-(\epsilon/r)^2$. For $z>1$, the velocity is suppressed at small $r$ by the factor $r^{2-2z}$, indicating that the particle takes longer to approach the deep interior. Now, solving the differential equation for $r(t)$ with the initial condition $r(0) = \epsilon$, we obtain
\begin{equation}\label{eq:lifshitz_geodesic}
r(t)=\left( \epsilon^{2z}+z^2 t^2 \right)^{\frac{1}{2z}}.
\end{equation}
Again, for $z>1$, the growth of $r(t)$ is slower at early times due to the exponent $1/(2z)$. 

\begin{figure}[h!]
	\begin{center}
		\includegraphics[scale=0.78]{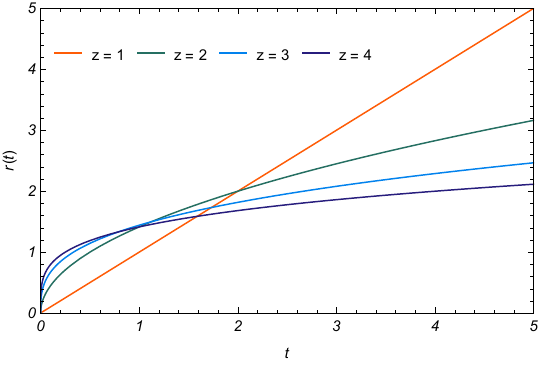}
			\end{center}
	\caption{The trajectory $r(t)$ of the infalling particle in the Lifshitz geometry for several values of the dynamical exponent $z$, with $\epsilon = 0.01$. For larger $z$, the particle takes longer to reach a given radial coordinate. The slope of the curves indicates power-law decay of the velocity for $z > 1$, in contrast to the constant asymptotic velocity observed for $z = 1$.}
	\label{fig:rtLif}
\end{figure}

These results reveal several key features of the geodesic motion in the Lifshitz background. For $z = 1$, which corresponds to the AdS case, the radial coordinate grows linearly at late times, $r(t) \sim t$, and the velocity approaches a constant, $\dot{r}(t) \to 1$. For $z > 1$, the late-time behavior of the radial coordinate is given by $r(t) \sim t^{1/z}$, which implies that the velocity scales as $\dot{r}(t) \sim t^{1/z - 1}$. Since $1/z - 1 < 0$ for $z > 1$, the velocity decays as a power law at late times, reflecting the slower infall of the particle into the interior compared to the relativistic case. The decay exponent becomes more negative as $z$ increases, indicating that larger values of the dynamical exponent lead to a more significant suppression of the radial velocity. The regime $z < 1$ is forbidden by the null energy condition and is therefore not considered. These features are summarized in Figure~\ref{fig:rtLif}.

Before proceeding to the computation of Krylov complexity, it is important to address potential concerns regarding the applicability of the holographic prescription in Lifshitz backgrounds. Lifshitz spacetimes exhibit several pathologies that distinguish them from the well-behaved AdS case, including null curvature singularities, trapping of non-radial null geodesics, degeneracy of causal wedges, and, in certain string theory embeddings, closed timelike curves \cite{Taylor:2015tka,Horowitz2012,Gentle2016,Copsey2011}. These features indicate that the standard holographic dictionary requires substantial revision for non-relativistic geometries. However, the specific nature of our calculation circumvents these obstructions. We consider only purely radial motion with no transverse momentum, which avoids the trapping of geodesics and the associated causal wedge degeneracy. The proper momentum prescription for Krylov complexity depends only on the local geometry along the radial geodesic and remains well-defined regardless of global causal pathologies. Thus, while the reconstruction of general bulk surfaces from boundary data may be obstructed in Lifshitz backgrounds, the momentum-Krylov correspondence remains applicable for the radial probes considered in this work.

\subsection{Generalised proper momentum and complexity}

Having established the trajectory and geodesic motion of a massive particle in the Lifshitz background, we now study the scaling and evolution of Krylov complexity using the prescription in Eq.~\eqref{CK}. To this end, we first compute two relevant radial momenta: the canonical momentum $P_r$ conjugate to the original radial coordinate $r$, and the proper momentum $P_\rho$ conjugate to the proper radial distance $\rho$. Recall that the latter is directly identified with the growth rate of Krylov complexity. 

Using the Lagrangian defined in Eq.~\eqref{eq:action_lifshitz}, the radial momentum conjugate to $r$ is given by
\begin{equation}
P_r=\frac{\partial \mathcal{L}}{\partial \dot{r}} = \frac{m \, r^{z-2} \dot{r}}{\sqrt{1 - r^{2z-2} \dot{r}^2}}. \label{eq:Pz_def}
\end{equation}
Substituting the velocity $\dot{r}(t)$ from Eq.~\eqref{eq:zdot2_lifshitz} into the above expression, we obtain
\begin{equation}
P_r=\frac{m}{r}\sqrt{\left( \frac{r}{\epsilon} \right)^{2z}-1}. \label{eq:Pz_finalLif}
\end{equation}
Having obtained $P_r$, we now examine its behavior in different regimes. Near the boundary, where $r \to \epsilon$ (early times), the square root vanishes, so $P_r \to 0$. This is consistent with the initial condition that the probe is released from rest at the UV cutoff. In the deep interior, $r \to \infty$ (late times), the square root tends to $1$, and the prefactor scales as $r^{z-1}$. For $z = 1$, $P_r$ approaches the constant $m \epsilon^{-1}$. For $z > 1$, $P_r$ grows without bound as $r^{z-1}$, reflecting the increasing canonical momentum as the particle falls inward.

With the canonical momentum at hand, we now turn to the proper momentum $P_\rho$, which enters directly in the holographic prescription for Krylov complexity. Based on Eq. \eqref{eq:proper_coordinate} The proper radial coordinate \(\rho\) is defined by \(d\rho = r^{-1} dr\), which ensures that the radial part of the metric becomes \(d\rho^2\) in the new coordinate. The proper momentum is then
\begin{equation}
P_\rho = \frac{\partial \mathcal{L}}{\partial \dot{\rho}} = r \, P_r, \label{eq:Prho_defLif}
\end{equation}
where we have used $\frac{\partial r}{\partial \rho} = r$. Substituting Eq.~\eqref{eq:Pz_finalLif} into the above expression, we obtain the explicit form of the proper momentum
\begin{equation}
P_\rho = m \,\sqrt{\left( \frac{r}{\epsilon} \right)^{2z} - 1}. \label{eq:Prho_finalLif}
\end{equation}
To find the time dependence of this quantity, we employ the geodesic solution $r(t)$ from Eq.~\eqref{eq:lifshitz_geodesic}. Inserting this into Eq.~\eqref{eq:Prho_finalLif} and simplifying the corresponding expressions, we arrive at
\begin{equation}
P_\rho(t)=\frac{m z}{\epsilon^z} \, t. \label{eq:P_rho_time}
\end{equation}
Note that we have chosen the positive sign since the proper momentum increases with time as the particle falls inward. This result reduces to the known expression in the relativistic limit $z = 1$ reported in \cite{Caputa:2024sux}.

Having obtained the proper momentum, the Krylov complexity follows directly from the holographic prescription in Eq.~\eqref{eq:complexity_rate}. Integrating $\dot{C}_K(t) = -P_\rho(t)$ with the initial condition $C_K(0) = 0$ gives	
\begin{equation}\label{LifCK}
C_K(t) = \int_0^t P_\rho(t') dt' = \frac{m z}{2\epsilon^z} t^2.
\end{equation}	
Thus, for all values of the dynamical exponent $z$, Krylov complexity exhibits universal quadratic growth at zero temperature, with the $z$-dependence appearing only in the overall coefficient.

The growth rate $\dot{C}_K(t)\sim P_\rho(t)$ is linear in time, with a coefficient proportional to $z$, while the exponent of $t$ is independent of the dynamical exponent. Consequently, Krylov complexity scales quadratically, $C_K(t) \sim t^2$, for all $z$. Thus, the non-relativistic scaling symmetry does not constrain the functional form of complexity growth; its only effect is an overall $z$-dependent factor in the rate.

\subsection{Early-time quadratic growth in Lifshitz field theory}

Having studied holographic Krylov complexity in the presence of a non-trivial dynamical exponent, this subsection addresses the field-theory perspective to corroborate and derive the early-time quadratic growth of this measure. The analysis focuses on Lifshitz-type field theories at finite temperature, following the frameworks established in Refs.~\cite{Vasli:2023,Imani:2025etp}. Calculations are presented for both scalar and Dirac fields, highlighting the universal quadratic-in-time behaviour, as well as the key differences in the coefficients between bosonic and fermionic systems.

A brief review of the computation of Krylov complexity in field-theory literature is in order. It is defined as the average position on the Krylov chain,
\begin{equation}
K_{\mathcal{O}}(t) = \sum_{n=0}^{\infty} n |\phi_n(t)|^2,
\label{eq:K_complexity}
\end{equation}
where the amplitudes $\phi_n(t)$ satisfy the discrete Schr\"odinger equation
\begin{equation}
i\frac{d\phi_n}{dt} = a_n\phi_n + b_{n+1}\phi_{n+1} + b_n\phi_{n-1},
\label{eq:schrodinger}
\end{equation}
with $\phi_{-1}(t) \equiv 0$ and initial condition $\phi_0(0)=1$. The Lanczos coefficients $\{a_n, b_n\}$ encode all information about the operator growth dynamics. The coefficients $b_n$ are obtained from the moments via the recursion relation
\begin{equation}
b_n = \sqrt{M_{2n}^{(n)}}, \qquad M_{2\ell}^{(j)} = \frac{M_{2\ell}^{(j-1)}}{b_{j-1}^2} - \frac{M_{2\ell-2}^{(j-2)}}{b_{j-2}^2},
\label{eq:lanczos_recursion}
\end{equation}
with $M_{2\ell}^{(-1)} = 0$ and $M_{2\ell}^{(0)} = \mu_{2\ell}$. The moments $\{\mu_{2n}\}$ are the Taylor expansion coefficients of the autocorrelation function,
\begin{equation}\label{eq:mun}
\mu_{2n}=\frac{1}{i^{2n}}\frac{d^{2n}\phi_0(t)}{dt^{2n}}\bigg|_{t=0}.
\end{equation}
For a massless scalar field, the autocorrelation function $\phi_0(t)$ is given by~\cite{Vasli:2023}
\begin{equation}
\phi_0^{S}(t) = \frac{\zeta\left(\xi-1, \frac{1}{2}+\frac{it}{\beta} \right) + \zeta\left(\xi-1, \frac{1}{2} - \frac{it}{\beta}\right)}{2(2^{\xi-1} - 1)\zeta(\xi-1)},
\label{eq:autocorr_scalar}
\end{equation}
where the scaling parameter is defined as $\xi \equiv (d-1)/z$. For a massless Dirac field, the autocorrelation function is~\cite{Imani:2025etp}
\begin{equation}
\phi_0^{D}(t) = \left(\frac{4t^2}{\beta^2} + 1\right)^{-\xi/2} \cos\left(\xi \tan^{-1}\frac{2t}{\beta}\right).
\label{eq:autocorr_dirac}
\end{equation}
To extract the early-time behavior, the above expressions are expanded for small $t$. For the scalar case, using properties of the Hurwitz zeta function yields
\begin{equation}\label{eq:phi0_scalar_exp}
\phi_0^{S}(t) = 1 - \frac{\xi(2^{\xi+1} - 1)(\xi - 1) \, \zeta(\xi+1)}{(2^\xi - 2)\beta^2 \, \zeta(\xi-1)} \, t^2 + \mathcal{O}(t^4).
\end{equation}
For the Dirac case, expanding Eq.~\eqref{eq:autocorr_dirac} gives
\begin{equation}
\phi_0^{D}(t) = 1 - \frac{2\xi(\xi+1)}{\beta^2} \, t^2 + \mathcal{O}(t^4).
\label{eq:phi0_dirac_exp}
\end{equation}
A key observation is that both scalar and Dirac fields exhibit quadratic-in-time behavior in the autocorrelation function at early times, although the coefficient differs between the two cases. This difference originates from the distinct spectral properties: the scalar spectral function is symmetric and bosonic, whereas the Dirac spectral function is antisymmetric and reflects Fermi--Dirac statistics. Interestingly, from the above expressions and Eq.~\eqref{eq:mun}, we see that at early times only the second moment controls the time dependence of the autocorrelation function and hence the complexity. Moreover, from the second moment, we obtain the first Lanczos coefficient, i.e., $b_1=\sqrt{\mu_2}$. Combining Eqs.~\eqref{eq:mun} and \eqref{eq:phi0_scalar_exp}, for the scalar case we obtain
\begin{equation}
b_1^{S} = \frac{1}{\beta}\sqrt{\frac{\xi(2^{\xi+1} - 1)(\xi - 1) \, \zeta(\xi+1)}{(2^{\xi-1} - 1)\, \zeta(\xi-1)}}.
\label{eq:mu2_scalar}
\end{equation}
On the other hand, from Eq.~\eqref{eq:phi0_dirac_exp}, for the Dirac field we find
\begin{equation}
b_1^{D} = \frac{2}{\beta}\sqrt{\xi(\xi+1)}.
\label{eq:b1_dirac}
\end{equation}
Now, in order to find the early-time Krylov complexity, we solve the Schr\"odinger equation Eq.~\eqref{eq:schrodinger} perturbatively. The first-order solution is
\begin{equation}
\phi_1(t) = -i b_1 t + \mathcal{O}(t^3),
\end{equation}
and thus the Krylov complexity becomes
\begin{equation}
K_{\mathcal{O}}(t) = \sum_{n=0}^{\infty} n |\phi_n(t)|^2 = b_1^2 t^2 + \mathcal{O}(t^4).
\end{equation}
This demonstrates the universal quadratic-in-time growth of Krylov complexity at early times. The coefficient of $t^2$ is precisely $b_1^2$, which equals the second moment $\mu_2$ of the power spectrum. Therefore, the early-time growth is controlled entirely by the second moment of the spectral function. Substituting the exact expressions for $\mu_2$ into the above result yields
\begin{equation}
K_{\mathcal{O}}^{S}(t) = \frac{\xi(2^{\xi+1} - 1)(\xi - 1) \, \zeta(\xi+1)}{(2^{\xi-1} - 1)\, \zeta(\xi-1)\beta^2} t^2 + \mathcal{O}(t^4),
\label{eq:final_result_scalar}
\end{equation}
and
\begin{equation}
K_{\mathcal{O}}^{D}(t) = \frac{4(d-1)(z+d-1)}{z^2\beta^2} t^2 + \mathcal{O}(t^4).
\label{eq:final_result_dirac}
\end{equation}
The key features of these results are as follows. First, both scalar and Dirac fields exhibit the universal quadratic growth $K(t) \propto t^2$ at early times. This is a robust feature of the Krylov complexity dynamics, arising from the perturbative solution of the Schr\"odinger equation. Second, while the $t^2$ growth is universal in form, the coefficients differ, with the Dirac coefficient being larger than the scalar one. Third, this difference has a clear physical origin: it arises from the fermionic nature of the Dirac field. The antisymmetric spectral function and the Fermi--Dirac distribution lead to a larger second moment, which in turn enhances the early-time complexity growth.

One can also extend this analysis to the massive case. In general, the autocorrelation function has a complicated exact expression for arbitrary dimension and dynamical exponent. For simplicity, let us focus on the scalar case with $d=5$ and $z=2$. In this case, the autocorrelation function is~\cite{Vasli:2023}
\begin{equation}
\phi_0^{S}(t) = \frac{\beta^2 \cos(m^2 t) - 2\beta t \sin(m^2 t)}{\beta^2 + 4t^2}.
\end{equation}
Expanding for early times gives
\begin{equation}
\phi_0^{S}(t) = 1 - \left(8+\beta m^2(4+\beta m^2)\right) \frac{t^2}{2\beta^2} + \mathcal{O}(t^4),
\end{equation}
so that the first Lanczos coefficient becomes
\begin{equation}
b_1^{S} = \frac{1}{\beta}\sqrt{8+\beta m^2(4+\beta m^2)}.
\end{equation}
The early-time Krylov complexity for the massive scalar is then
\begin{equation}
K_{\mathcal{O}}^{S}(t) = \left(8+\beta m^2(4+\beta m^2)\right) \frac{t^2}{\beta^2}+ \mathcal{O}(t^4).
\label{eq:final_result_massive_scalar}
\end{equation}
Thus, the quadratic growth persists, but the mass adds an additional contribution to the coefficient. The above analysis can be straightforwardly extended to other values of $d$ and $z$. 

For the massive Dirac field, the diagonal Lanczos coefficients are non-zero, $a_n = \mp (-1)^n m^z$, which introduces additional structure and richness in the Krylov evolution~\cite{Imani:2025etp}. The off-diagonal coefficients $b_n$ remain as in the massless case, but the presence of $a_n$ modifies the dynamics at higher orders. Although the computations become more involved, the early-time quadratic growth remains qualitatively the same.

Our derivation reveals two key aspects of Krylov complexity growth in Lifshitz field theories. First, the early-time growth is universally quadratic in time, $K_{\mathcal{O}}(t) = b_1^2 t^2 + \mathcal{O}(t^4)$, regardless of the field statistics. This universality stems from the fundamental structure of the Krylov construction: the first Lanczos coefficient $b_1$ is always determined by the second moment of the spectral function, and the perturbative solution of the Schr\"odinger equation always yields $|\phi_1(t)|^2 \propto t^2$ at leading order. Second, despite the universal $t^2$ form, the coefficient $b_1^2 = \mu_2$ is theory-dependent. The Dirac coefficient exceeds the scalar one, and this enhanced growth can be understood physically: the fermionic spectral function is antisymmetric, leading to a larger spread in the frequency domain; the Fermi--Dirac statistics introduce additional correlations; and the cosine factor in the Dirac autocorrelation function contributes an extra $t^2$ term from its expansion. These features reflect the fundamental distinction between bosonic and fermionic quantum fields, which persists even in the non-relativistic Lifshitz setting. These results provide a rigorous field-theoretic foundation for understanding Krylov complexity dynamics in non-relativistic systems with Lifshitz scaling, while clarifying the distinctions between bosonic and fermionic statistics that become apparent at the level of the growth coefficients.

\section{Hyperscaling-violating generalization}\label{hvback}

We now generalize the previous analysis to include the hyperscaling-violating exponent $\theta$. As we will see, the presence of $\theta$ introduces new scaling dimensions that modify both the geodesic motion and the growth of Krylov complexity. We begin by deriving the particle trajectory in the hyperscaling-violating background and computing the associated proper momentum. For generic values of the parameters, the time evolution of the proper momentum and complexity cannot be expressed in closed analytic form; however, the asymptotic behavior can be extracted systematically, which we present in the following subsections.

\subsection{Geodesic motion and complexity}

The metric in Poincar\'e-like coordinates is given by Eq.~\eqref{eq:metricNR}. We consider a radial geodesic with $\vec{x} = \text{constant}$, which we set to zero by translation invariance. The particle is initially released from rest at $r(0) = \epsilon$, so that $\dot{r}(0) = 0$. For a trajectory with $d\vec{x} = 0$, the metric reduces to
\begin{equation}
ds^2 = - r^{-2z + 2\theta/d} dt^2 + r^{-2 + 2\theta/d} dr^2. \label{eq:hsv_radial_metric}
\end{equation}
Using the proper time interval $d\tau^2 = -ds^2$, the action for a massive particle of mass $m$ takes the form
\begin{equation}
S = -m \int r^{-z + \theta/d} \sqrt{1 - r^{2z-2} \dot{r}^2} \, dt. \label{eq:hsv_action}
\end{equation}
Since the Lagrangian $\mathcal{L}(r,\dot{r})$ has no explicit time dependence, the associated Hamiltonian (energy) is conserved. Denoting this constant by $E$, a straightforward calculation yields
\begin{equation}
E = \frac{m \, r^{-z + \theta/d}}{\sqrt{1 - r^{2z-2} \dot{r}^2}} = m \, \epsilon^{-z + \theta/d}, \label{eq:hsv_energy}
\end{equation}
where the second equality follows from evaluating the expression at the initial position $r = \epsilon$ with $\dot{r} = 0$. Solving Eq.~\eqref{eq:hsv_energy} for the radial velocity squared gives
\begin{equation}
\dot{r}^2 = r^{2-2z} \left( 1 - \left( \frac{\epsilon}{r} \right)^{2z - 2\theta/d} \right). \label{eq:hsv_zdot}
\end{equation}
In what follows, we choose the positive square root, corresponding to the particle falling inward from the UV boundary ($\dot{r} > 0$ for $t > 0$). Integrating the above equation and employing the initial conditions, we find
\begin{equation}
t = \frac{r^{z}}{z} \, {}_2F_1\!\left(\frac{1}{2}, \frac{d z}{2\theta - 2d z}; 1 + \frac{d z}{2\theta - 2d z}; \left(\frac{r}{\epsilon}\right)^{\frac{2\theta}{d} - 2z}\right) + \epsilon^{z} \frac{d \sqrt{\pi} \, \Gamma\!\left(\frac{d z}{2\theta - 2d z}\right)}{(\theta - 2d z) \, \Gamma\!\left(-1 + \frac{\theta}{2\theta - 2d z}\right)}, \label{eq:hsv_geodesic_time}
\end{equation}
where \({}_2F_1\) is the hypergeometric function. In general, it is not possible to find an explicit closed-form expression for $r(t)$, and a numerical illustration is required.

Let us now compute the two relevant radial momenta: the canonical momentum $P_r$ conjugate to the coordinate $r$, and the proper momentum $P_\rho$ conjugate to the proper radial distance $\rho$. The canonical momentum is given by
\begin{equation}
P_r = \frac{\partial \mathcal{L}}{\partial \dot{r}} = \frac{m \, r^{z + \theta/d - 2} \dot{r}}{\sqrt{1 - r^{2z-2} \dot{r}^2}}. \label{eq:hsv_conjugate_momentum}
\end{equation}
Substituting the velocity expression from Eq.~\eqref{eq:hsv_zdot} into the above equation, we obtain
\begin{equation}
P_r = m \, \epsilon^{-z + \theta/d} \, r^{z-1} \sqrt{1 - \left( \frac{\epsilon}{r} \right)^{2z - 2\theta/d}}. \label{eq:Pz_final_hsv}
\end{equation}
With the expression for $P_r$ at hand, we now explore its behavior across different regimes. Near the boundary the square root vanishes and hence $P_r \to 0$, which is consistent with the probe starting from rest at the UV cutoff. In the opposite limit, deep in the interior, the square root approaches unity and the prefactor scales as $r^{z-1}$. Interestingly, this power-law growth is independent of $\theta$, implying that the late-time divergence of $P_r$ is dictated solely by the dynamical exponent $z$. The hyperscaling-violating exponent $\theta$ only enters through the normalization factor $\epsilon^{-z + \theta/d}$. For $z = 1$, $P_r$ tends to a finite constant $m \epsilon^{-1 + \theta/d}$, which differs from the standard AdS result due to the presence of $\theta$. For $z > 1$, $P_r$ diverges as $r^{z-1}$, reflecting the steady growth of canonical momentum as the particle falls deeper into the bulk. In contrast to the Lifshitz case, where the normalization was $\epsilon^{-z}$, here the presence of $\theta$ modifies the effective UV scale without altering the IR scaling behavior.

The proper radial coordinate $\rho$ is defined by $d\rho = r^{\theta/d - 1} dr$, which ensures that the radial part of the metric becomes $d\rho^2$ in the new coordinate. The proper momentum is then given by $P_\rho = m \, d\rho/d\tau$, and a straightforward computation yields
\begin{equation}
P_\rho = m  \,  \sqrt{\left(\frac{r}{\epsilon}\right)^{2z - 2\theta/d}-1}. \label{eq:Prho_final_hsv}
\end{equation}
Note that the two momenta are related by $P_\rho = r^{1 - \theta/d} P_r$. For $\theta = 0$, this relation reduces to the Lifshitz result $P_\rho = r P_r$, as expected. In general, it is not possible to find the time dependence of complexity analytically, but an asymptotic expansion is possible, as we carry out in the next section.

\begin{figure}[h!]
	\begin{center}
		\includegraphics[scale=0.78]{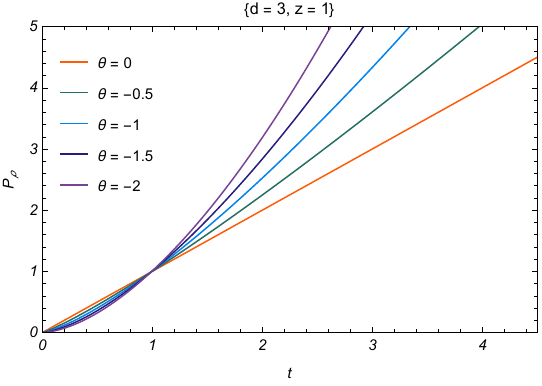}
		\hspace*{1cm}
				\includegraphics[scale=0.78]{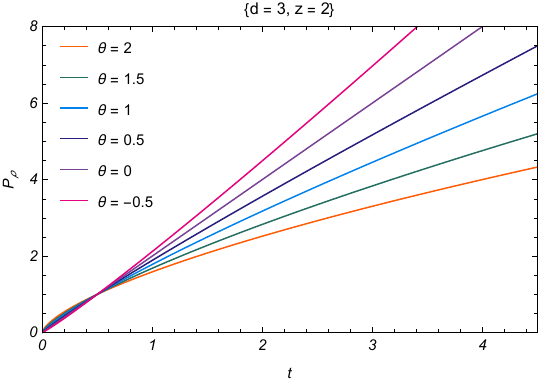}
			\end{center}
	\caption{Proper momentum of a massive particle for several values of the dynamical and hyperscaling-violating exponents in $d=3$, with $\epsilon = 0.01$.}
	\label{fig:prho}
\end{figure}

To conclude this section, we examine the explicit dependence of the complexity growth rate — which is equal to the proper momentum — on the dynamical and hyperscaling-violating exponents, as shown in Figure~\ref{fig:prho} for the representative case $d=3$. The qualitative features are largely independent of the spacetime dimensionality, so this choice is sufficient. For clarity, we have scaled the proper momentum by $m \, \epsilon^{-z + \theta/d}$. Several observations are in order. First, although the rate of holographic complexity growth decreases with increasing $z$ and $\theta$ at early times, it increases with these parameters at sufficiently late times. The time scale at which this crossover occurs depends on the dynamical exponent, decreasing for larger values of $z$. Second, consistent with our earlier observations, exact linear growth — corresponding to quadratic complexity — occurs only in the Lifshitz limit ($\theta = 0$), while for non-zero $\theta$, the complexity clearly exhibits non-quadratic growth. Finally, as seen from the right panel, the complexity evolution for the case $\theta = d-1$, which corresponds to a boundary theory with a Fermi surface, does not differ qualitatively from the other cases considered. This is noteworthy because, in this case, the scaling of holographic entanglement entropy is significantly affected and exhibits a violation of the area law \cite{Ogawa2012}. Our results suggest that holographic Krylov complexity is largely insensitive to this particular feature of the entanglement structure. In this sense, this measure appears to be a less sensitive probe of the underlying physics of the boundary theory.

\subsection{Asymptotic expansions}

In this section, we determine the early-time and late-time scaling of Krylov complexity for generic values of the Lifshitz and hyperscaling-violating exponents. Using Eq.~\eqref{eq:hsv_geodesic_time}, we analyze the behavior of the geodesic solution in the relevant limits.

At early times, corresponding to $r \to \epsilon$, we obtain the expansion
\begin{equation}
t = \frac{\sqrt{2} \, \epsilon^{z - 1/2} \sqrt{r - \epsilon}}{\sqrt{z - \theta/d}} \left(1 + \frac{3d(2z - 1) - 2\theta}{12 d \, \epsilon} (r - \epsilon)\right) + \cdots. \label{eq:t_early_hsv}
\end{equation}
Inverting this relation to express $r$ in terms of $t$, we find
\begin{equation}
r = \epsilon \left(1 + \frac{d z - \theta}{2d} \left(\frac{t}{\epsilon^z}\right)^2 + \frac{(2\theta + 3d(1-2z))(d z - \theta)^2}{24 d^3} \left(\frac{t}{\epsilon^z}\right)^4\right) + \cdots. \label{eq:r_early_hsv}
\end{equation}
These expansions capture the leading corrections to the initial position as the particle begins to fall inward. For $\theta = 0$, they reduce to the Lifshitz results derived in the previous section, with the first correction scaling as $t^2$ and a coefficient that depends on both $z$ and $\theta$.

Using the expression for the canonical momentum in Eq.~\eqref{eq:Pz_final_hsv}, together with the early-time expansion for $r(t)$, we obtain
\begin{equation}
P_r = \frac{m (d z - \theta)}{d \, \epsilon^{1 - \theta/d}} \frac{t}{\epsilon^z} \left(1 - \frac{(3d - 2\theta)(d z - \theta)}{6 d^2} \left(\frac{t}{\epsilon^z}\right)^2\right) + \cdots. \label{eq:Pz_early_hsv}
\end{equation}
Similarly, the proper momentum $P_\rho$ from Eq.~\eqref{eq:Prho_final_hsv} gives
\begin{equation}
P_\rho = \frac{m (d z - \theta)}{d} \frac{t}{\epsilon^z} \left(1 - \frac{\theta (d z - \theta)}{6 d^2} \left(\frac{t}{\epsilon^z}\right)^2\right) + \cdots. \label{eq:Prho_early_hsv}
\end{equation}
These expressions reveal that both momenta grow linearly in time at leading order, with coefficients proportional to $(d z - \theta)/d$, which vanishes when $\theta = d z$, signaling a special limiting case. In the relativistic limit $(z = 1, \theta = 0)$, the above expansions reduce to the familiar results
\begin{equation}
P_r = \frac{m t}{\epsilon^2} - \frac{m t^3}{2 \epsilon^4} + \cdots, \qquad P_\rho = \frac{m t}{\epsilon} + \cdots. \label{eq:relativistic_limit_early}
\end{equation}
Thus, at early times, the hyperscaling-violating exponent modifies the amplitude of the linear growth while preserving its functional form. 

We now turn to the late-time behavior, which exhibits distinct scaling regimes depending on the value of $\theta$ relative to $d z$. Starting from Eq.~\eqref{eq:hsv_geodesic_time}, at late times $r \to \infty$ (assuming $\frac{2\theta}{d} - z < 0$ for convergence of the hypergeometric series), we obtain
\begin{equation}
r = (z t)^{1/z} \left(1 + \frac{d}{2(d z - 2\theta)} \left(\frac{z t}{\epsilon^z}\right)^{\frac{2\theta}{d z} - 2}\right) + \cdots. \label{eq:r_late_hsv}
\end{equation}
This asymptotic form captures the leading power-law growth of the radial coordinate, with subleading corrections controlled by the combination $d z - 2\theta$. For $\theta = 0$, this reduces to the Lifshitz result $r \sim (z t)^{1/z}$. Substituting the above expansion into the expressions for the canonical and proper momenta, we find the late-time scaling
\begin{align}
P_r &= m \, \epsilon^{\theta/d - z} (z t)^{1 - 1/z}, \label{eq:Pz_late_hsv} \\
P_\rho &= m \, \epsilon^{\theta/d - z} (z t)^{1 - \theta/(d z)}. \label{eq:Prho_late_hsv}
\end{align}
These results reveal several important features. First, the canonical momentum $P_r$ scales as $t^{1 - 1/z}$, which is independent of $\theta$ and matches the Lifshitz result. In contrast, the proper momentum $P_\rho$ exhibits a modified exponent $1 - \theta/(d z)$, which explicitly depends on the hyperscaling-violating exponent. For $\theta > 0$, the growth of $P_\rho$ is suppressed compared to the Lifshitz case, while for $\theta < 0$, it is enhanced. In the relativistic limit $(z = 1, \theta = 0)$, the above expressions reduce to
\begin{equation}
P_r = \frac{m}{\epsilon} + \cdots, \qquad P_\rho = \frac{m t}{\epsilon} + \cdots. \label{eq:relativistic_limit_late}
\end{equation}
Thus, the relativistic limit recovers the expected constant canonical momentum and linearly growing proper momentum, in agreement with known results in the literature.

The corresponding complexity growth follows from integration of the proper momentum. At early times, Eq.~\eqref{eq:Prho_early_hsv} yields	
\begin{equation}\label{CKearly}
C_K(t)=\frac{m(dz-\theta)}{2d}\,\frac{t^2}{\epsilon^z} + \mathcal{O}(t^4),
\end{equation}	
which exhibits the expected quadratic growth. Interestingly, for the special case $\theta = d z$, the null energy condition restricts the dynamical exponent to the interval $1 \leq z \leq \frac{d}{d-1}$, and the above expansion is no longer valid. In this case, a separate treatment is required. Indeed, the action simplifies considerably and the conserved energy becomes $E = m$. As can be seen from Eq.~\eqref{eq:hsv_zdot}, the radial velocity squared vanishes identically, and after imposing the initial condition we find that the particle remains at a fixed radial position. Consequently, the proper momentum is identically zero, which implies that the Krylov complexity growth rate vanishes, $\dot{C}_K(t) = 0$, and the complexity remains constant at all times. This case corresponds to a trivial geodesic where the probe does not fall inward, and thus no complexity growth occurs. This behavior is consistent with the structure of the geodesic equation: the relevant Christoffel symbols appearing in the radial probe equation, such as $\Gamma^r_{tt}$ and $\Gamma^t_{rt}$, are proportional to $(\theta-dz)$ and therefore vanish identically when $\theta = d z$. Hence, the gravitational pull on the infalling probe disappears, leading to a static solution.%
\footnote{As a side comment, we note that, as argued in Ref.~\cite{Kachru:2008yh}, in this non-relativistic geometry an object of fixed proper energy $E_{\text{pr}}$ redshifts according to $E(r) = E_{\text{pr}} / r^{z - \theta/d}$. Thus, in the specific limit $\theta = d z$, the energy becomes observer-independent. Although we do not yet fully understand the physical interpretation behind the relation between a trivial complexity and this behavior, we find the observation suggestive and worth mentioning.}

On the other hand, the late-time scaling from Eq.~\eqref{eq:Prho_late_hsv} gives	
\begin{equation}\label{CKlate}
C_K(t) \sim \frac{m\,\epsilon^{\theta/d-z}}{2z - \theta/d}\, (zt)^{2 - \theta/(dz)}.
\end{equation}	
This explicitly shows that the hyperscaling-violating exponent modifies the power law of complexity growth at late times.
The expansion in the above equation breaks down when $\theta = 2d z$, as the prefactor $2 - \theta/(d z)$ vanishes. Again, in this special case, a separate treatment is required. The action simplifies considerably, and the resulting Lagrangian leads to a conserved energy $E = m \epsilon^z$. The radial velocity squared then takes the form
\begin{equation}
\dot{r}^2 = r^{2-2z} \left( 1 - \left( \frac{r}{\epsilon} \right)^{2z} \right).
\end{equation}
Integrating this equation and imposing the initial condition $r(0) = \epsilon$, we find
\begin{equation}
r(t) = \epsilon \left( \cos \frac{z t}{\epsilon^z} \right)^{1/z}.
\end{equation}
In this case, the proper momentum is given by $P_\rho = -m \tan \frac{z t}{\epsilon^z}$ which yields a Krylov complexity of the form
\begin{equation}\label{log}
C_K(t) \sim \frac{m \, \epsilon^z}{z} \, \log \left( \cos \frac{z t}{\epsilon^z} \right).
\end{equation}
This expression exhibits a logarithmic dependence on the cosine, giving rise to oscillatory behavior. Consequently, the complexity is not simply logarithmic in time, but rather oscillatory with a logarithmic envelope. The above result reduces to the early-time expression in Eq.~\eqref{CKearly} and displays oscillatory behavior at late times, with periodic divergences at $t = (\pi/2)(\epsilon^z/z)$. This suggests that the present case is qualitatively distinct from the generic hyperscaling-violating regime and may indicate a transition to a different phase of complexity growth. It is worth noting that the null energy condition imposes nontrivial constraints on the allowed parameter space. For the special case $\theta = 2d z$, the NEC requires $\frac{d}{2d-1} \leq z \leq 1$. Hence, the logarithmic and oscillatory behavior described above is realized only within this restricted range of the dynamical exponent. Furthermore, in the early-time limit, the leading scaling of Eq.~\eqref{log} reduces to $C_K(t)\sim t^2$, which is consistent with Eq.~\eqref{CKearly}.

\section{Top-down solutions}\label{topdown}

The bottom-up analysis of the previous sections captures the essential scaling properties of Lifshitz and hyperscaling-violating geometries, but its phenomenological nature leaves open the question of whether such backgrounds can be embedded in a UV-complete theory. In this section, we address this by considering explicit top-down constructions derived from string theory and supergravity. We examine the geodesic motion and Krylov complexity in these backgrounds and compare the results with the bottom-up findings.

\subsection{Lifshitz fixed points from string theory}

In this subsection, we review a top-down string-theory construction of Lifshitz-like fixed points in type IIB supergravity, originally presented in \cite{Azeyanagi:2009pr}. This provides a concrete holographic realisation of anisotropic scale invariance with a known microscopic interpretation, serving as a natural setting to extend our analysis of Krylov complexity to non-relativistic backgrounds.

The type IIB supergravity solutions are expected to be dual to certain Lifshitz-like fixed points with anisotropic scale invariance, describing a class of D3--D7 systems. In particular, there exist solutions that interpolate between these anisotropic fixed points in the IR and the standard AdS$_5$ geometry in the UV. These RG flows are triggered by a non-zero theta-angle in the dual Yang--Mills theory that depends linearly on one of the spatial coordinates.

Starting from the type IIB supergravity action in the string frame and assuming an ansatz that preserves three-dimensional Lorentz symmetry $SO(2,1)$, the authors find a scaling solution in which all metric functions are logarithmic in the radial coordinate $r$. In the Einstein frame, the solution takes the form\footnote{For simplicity, we only report the metric here. The full relation between the action parameters and the solution can be found in \cite{Azeyanagi:2009pr}.}
\begin{equation}
ds_E^2 = \frac{11}{12} R^2 \left[ r^2 (-dt^2 + dx^2 + dy^2) + r^{4/3} dw^2 + \frac{dr^2}{r^2} \right] + R^2 ds_{X_5}^2.
\end{equation}
The above metric is invariant under the anisotropic scaling transformation
\begin{equation}
(t, x, y, w, r) \to \left( \lambda t, \lambda x, \lambda y, \lambda^{2/3} w, \frac{r}{\lambda} \right),
\end{equation}
which corresponds to a Lifshitz-like fixed point with dynamical exponent $z = 3/2$ for the $(x,y)$ directions. The anisotropic direction $w$ has a different scaling dimension from the $(x,y)$ directions, reflecting the breaking of rotational symmetry. The dual field theory corresponds to a deformation of $\mathcal{N}=4$ super Yang--Mills theory on $\mathbb{R}^{1,3}$ triggered by a $\Theta$-angle that depends linearly on the $w$ coordinate, e.g., $\Theta(w) = \frac{2\pi}{L} w$, which appears in the Lagrangian as $\delta \mathcal{L} = \frac{\Theta(w)}{8\pi^2} \, \text{Tr}(F \wedge F)$. When the $w$-direction is compactified, this becomes a three-dimensional Chern--Simons term. The solution exhibits an RG flow from $AdS_5 \times X_5$ in the UV to the anisotropic scaling solution in the IR.

Now, in order to study Krylov complexity in this background, we consider a massive particle moving along a purely radial geodesic, with all spatial coordinates held constant ($dx = dy = dw = 0$). Under this condition, because the anisotropy parameter is encoded in the coefficient of $dw$, the induced metric on the particle trajectory reduces to the corresponding one in the asymptotically AdS geometry. Consequently, all the results for the proper momentum and the time dependence of complexity are identical to the relativistic case. In particular, a straightforward computation — which we omit as it is entirely trivial — shows that the complexity grows quadratically in time, in accordance with the previous results. This is also consistent with the bottom-up approach with only a Lifshitz dynamical exponent, reported in Eq.~\eqref{LifCK}, which shows that the dynamical exponent alone does not affect the complexity growth rate.

This top-down construction provides several features that make it ideal for extending our holographic analysis of Krylov complexity to non-relativistic backgrounds. The solution has a known D-brane interpretation and controlled supergravity description, unlike phenomenological Lifshitz metrics. The anisotropic scaling of the $(x,y)$ and $w$ directions allows us to study direction-dependent complexity growth. The non-trivial dilaton $e^{\phi} = r^{2/3}$ introduces new couplings that may affect the probe action and consequently the complexity dynamics. Finally, the compact internal space $X_5$ opens the possibility of studying extended probes and symmetry-resolved complexity, along the lines of Ref.~\cite{Nastase:2026lhz}.

\subsection{Top-down hyperscaling violation from intersecting branes}

Non-relativistic symmetries in field theory and gravity fall into two main classes: Schr\"odinger and Lifshitz symmetries. Both involve anisotropic scaling between time and space, breaking Lorentz invariance, but differ significantly in algebraic structure and embeddability in string theory. Schr\"odinger symmetry is a rich extension of the Galilei algebra, including dilatation, special conformal transformations, and a $U(1)$ particle-number symmetry, making it well-suited for describing systems with fixed particle density, such as fermions at unitarity. Schr\"odinger spacetimes admit clean string-theory embeddings. Lifshitz symmetry, by contrast, is minimal, retaining only translations, rotations, and anisotropic scaling, with no boosts or particle-number conservation. While powerful for modelling quantum critical points, Lifshitz geometries have proven difficult to embed in string/M-theory due to challenges in matching the required scaling with supergravity equations \cite{Hartnoll:2009ns}.

The paper \cite{Dey:2012tg} provides a systematic construction of Lifshitz-like spacetimes with hyperscaling violation from intersecting brane configurations in string/M-theory. Starting from the well-known $1/4$ BPS threshold D1-D5 bound state of type IIB string theory, the authors apply successive S- and T-duality transformations to generate a family of F-Dp bound states with $0 \leq p \leq 5$. These solutions are distinct from the more familiar $1/2$ BPS non-threshold F-Dp bound states and preserve supersymmetry throughout the construction. The duality chain proceeds from D1-D5 through two T-dualities to yield D0-D4 and subsequently D1-D3, followed by an S-duality transformation to produce F-D3, and then further T-dualities along D3-brane directions to obtain F-D2, F-D1, and F-D0, while T-dualities along common transverse directions produce F-D4 and F-D5. Being U-dual to the D1-D5 system, all these solutions are $1/4$ BPS threshold bound states. Interestingly, for $p = 4$, the familiar Lifshitz-like behavior is absent. In this special case, the near-horizon limit collapses to a simple $AdS_2$ space.

The near-horizon limits of these intersecting F-Dp solutions yield Lifshitz-like spacetimes characterised by a dynamical critical exponent $z$ and a hyperscaling violation exponent $\theta$. The general metric in the string frame takes the form
\begin{equation}
ds^2 = \frac{Q_2^{\frac{2}{4-p}}}{r^{\frac{4(3-p)}{4-p}}}\left(-\frac{dt^2}{Q_1 Q_2 r^{\frac{2(6-p)}{4-p}}} + \frac{\sum_{i=1}^p (dx^i)^2}{Q_2}+ \frac{(dx^{p+1})^2}{Q_1} + \frac{4}{(4-p)^2} dr^2 +r^2 d\Omega_{7-p}^2\right),
\end{equation}
with the dilaton field and the NS-NS and R-R form fields also given explicitly in the paper. Here $ Q_1$ and $Q_2$ are the charges associated with D1 and D5 branes.  This metric represents the complete supergravity solution in the string frame and is a genuine string theory background obtained from intersecting branes. The sphere $S^{7-p}$ represents the internal compact directions, and the dilaton is generally non-constant except for $p=1$. This full 10-dimensional metric is essential because it shows that the Lifshitz-like geometry emerges from a genuine top-down string/M-theory setup.

To extract the dual field theory properties, the authors compactify on the $S^{7-p}$ sphere and transform to the Einstein frame. This yields the reduced $(p+3)$-dimensional metric\footnote{We follow the formulation of ~\cite{Dey:2012tg}, but with the radial coordinate denoted as $r$ instead of $u$ to maintain consistency with the notation used throughout this paper.}
\begin{equation}
ds_{p+3}^2 = \frac{Q_1^{\frac{2}{p+1}} Q_2}{r^{\frac{4}{(4-p)(p+1)}}} \left(-\frac{dt^2}{Q_1 Q_2 r^{\frac{2(6-p)}{4-p}}} + \frac{\sum_{i=1}^{p} (dx^i)^2}{Q_2} + \frac{(dx^{p+1})^2}{Q_1} + \frac{4}{(4-p)^2}dr^2 \right),
\end{equation}
where the sphere directions have been integrated out. This metric describes the effective geometry where the dual field theory lives, with the boundary of this spacetime being where the field theory is defined. This reduced Einstein frame metric is essential because it reveals the field theory interpretation through the critical exponents, whereas the sphere directions and the string frame would obscure the field-theoretic scaling. The non-relativistic scaling properties of the dual field theory are encoded in the critical exponents
\begin{equation}\label{ztheta}
z = \frac{2(p-5)}{p-4}, \qquad \theta = p + \frac{p-2}{p-4},
\end{equation}
where $d = p+1$ is the spatial dimension of the boundary theory. The dilaton is generally non-constant except for the $p=1$ case, generating holographic RG flows with distinct phase structures in different parameter regimes. The validity of the gravity description requires both the effective string coupling and spacetime curvature to remain small, leading to well-defined ranges of the radial coordinate where the supergravity approximation can be trusted, with the solutions requiring uplift to M-theory for type IIA or passage to the S-dual frame for type IIB outside these regions.

\subsubsection{Geodesic motion and Krylov complexity}

We derive the geodesic motion and proper momentum for a massive probe in the reduced $(p+3)$-dimensional Einstein frame metric obtained after compactification on $S^{7-p}$. This analysis provides the foundation for computing holographic Krylov complexity in the dual field theory.

We consider a massive particle of mass $m$ moving along a purely radial geodesic, with all spatial coordinates held constant ($dx^i = 0$ for all $i$). Under this condition, the metric reduces to
\begin{equation}
ds^2 = - Q_1^{\frac{2}{d}-1} \, r^{2\frac{\theta}{d} - 2z} dt^2 + \frac{4 Q_1^{\frac{2}{d}} Q_2}{(p-4)^2} r^{2\frac{\theta}{d} - 2} dr^2,
\end{equation}
For convenience, we define the constants $A \equiv Q_1^{\frac{1}{d}-\frac{1}{2}}$ and $B \equiv \frac{4 Q_1 Q_2}{(4-p)^2}$ and thus the action for the massive particle can be written as
\begin{equation}
S=-m A \int r^{\frac{\theta}{d}-z} \sqrt{1 - B r^{2z-2} \dot{r}^2} \, dt,
\end{equation}
where $\dot{r} \equiv dr/dt$. Since the Lagrangian has no explicit time dependence, the associated Hamiltonian (energy) is conserved. A straightforward computation gives
\begin{equation}
E = \frac{m A r^{\frac{\theta}{d}-z}}{\sqrt{1 - B r^{2z-2} \dot{r}^2}} = m A \epsilon^{\frac{\theta}{d}-z},
\end{equation}
where the second equality follows from the initial condition that the particle is released from rest at the UV cutoff. Solving for the velocity squared yields the radial geodesic equation
\begin{equation}\label{udot}
\dot{r}^2 = \frac{1}{Br^{2z-2}}  \left[ 1 - \left(\frac{r}{\epsilon}\right)^{2\frac{\theta}{d}-2z} \right].
\end{equation}
Now the canonical momentum conjugate to the radial coordinate $r$ is
\begin{equation}
P_r = m A \sqrt{B} \, \epsilon^{\frac{\theta}{d}-z} r^{z-1} \sqrt{1 - \left(\frac{r}{\epsilon}\right)^{2\frac{\theta}{d}-2z}}.
\end{equation}
This expression gives the canonical momentum as a function of the radial coordinate $r$. Using the proper radial distance defined by $d\rho^2 = A^2 B \, r^{2\theta/d - 2} \, dr^2$, and substituting the expression for $P_r$, the proper momentum is given by
\begin{equation}
P_\rho = m \sqrt{\left(\frac{r}{\epsilon}\right)^{2z - 2\theta/d} - 1}.
\end{equation}
This is the proper momentum that enters the holographic prescription for Krylov complexity.

\subsubsection{Time dependence of complexity}

According to the holographic proposal~\cite{Caputa:2024sux}, the growth rate of Krylov complexity is given by
$\dot{C}_K(t) = -P_\rho(t)$, therefore
\begin{equation}
C_K(t) = \int_0^t P_\rho(t') \, dt',
\end{equation}
with the initial condition $C_K(0) = 0$. To obtain the explicit time dependence, we need to invert the geodesic equation to find $r(t)$. From the geodesic equation,
\begin{equation}
t=\sqrt{B} \int_{\epsilon}^{r} \frac{u^{z-1}}{\sqrt{1 - \left(u/\epsilon\right)^{2\frac{\theta}{d}-2z}}} \, du.
\end{equation}
For general values of $\theta$ and $z$, this integral yields a hypergeometric function. However, the asymptotic behavior can be extracted analytically. Let us briefly discuss the early-time and late-time expansions, following the same procedure as in the previous sections.

\textbf{i) Early-time behavior ($r \to \epsilon$):}

Expanding the velocity near the cutoff and integrating with the initial condition $r(0)=\epsilon$, we obtain
\begin{equation}
r(t) \approx \epsilon \left[ 1 + \frac{dz-\theta}{2dB} \frac{t^2}{\epsilon^{2z}} \right].
\end{equation}
This shows that the particle departs from the boundary quadratically in time, with the coefficient controlled by the critical exponents and $B$, as well as the UV cutoff. Substituting this into the expression for the proper momentum gives the early-time behavior
\begin{equation}
P_\rho(t) \approx \frac{m(\theta-dz)}{d\sqrt{B}} \frac{t}{\epsilon^z}.
\end{equation}
Thus, the proper momentum grows linearly at early times, with a prefactor that encodes the combined effect of the hyperscaling-violating and dynamical exponents. Consequently, Krylov complexity exhibits quadratic growth at early times,
\begin{equation}
C_K(t) \approx \frac{m(dz-\theta)}{2d\sqrt{B}} \frac{t^2}{\epsilon^z},
\end{equation}
which is the expected universal early-time scaling for complexity, with non-trivial modifications arising from the non-relativistic background parameters. Moreover, we note that the above result coincides with the bottom-up analysis of the previous section, Eq.~\eqref{CKearly}, up to a constant factor. Hence, the time scaling of Krylov complexity in both approaches is the same, confirming that the early-time growth is a robust feature insensitive to the specific top-down embedding. It is worth noting that, in contrast to the previous section where the special case $\theta = dz$ caused the overall prefactor to vanish, here the exponents $z$ and $\theta$ depend on $p$ in such a way that for no value of $p$ do the coefficients vanish. Thus, the above result holds for all cases (see Table~\ref{tab:exponents}). As already discussed, the analysis is valid for $0<p<5$. Within this range, however, the case $p=4$ stands out that the near-horizon limit no longer yields a Lifshitz-like spacetime but instead simplifies to pure $AdS_2$.
\begin{table}[h]
	\centering
	\begin{tabular}{|c|c|c|c|}
		\hline
		$p=(d-1)$ & $z$ & $\theta$ & $C_K(t) \sim$ \\
		\hline
		0 & 5/2 & 1/2 & $t^{9/5}$ \\
		1 & 8/3 & 4/3 & $t^{7/4}$ \\
		2 & 3 & 2 & $t^{16/9}$ \\
		3 & 4 & 2 & $t^{15/8}$ \\
		5 & 0 & 8 & no growth \\
		\hline
	\end{tabular}
	\caption{Late-time growth exponents of $\mathcal{C}_K(t)$ for various $F-Dp$ bound states, with exponent $2 - \theta/dz$. The analysis holds for $0 < p < 5$, with $p = 4$ being special where the near-horizon geometry becomes AdS$_2$ rather than Lifshitz-like. At $p = 5$, $z = 0$, and the complexity shows no growth.}
	\label{tab:exponents}
\end{table}

\textbf{ii) Late-time behavior ($r \to \infty$):}

In the deep interior, where $r \gg \epsilon$, the square root in the geodesic equation approaches unity, and the radial velocity simplifies significantly. The resulting differential equation is easily integrated and yields the late-time trajectory
\begin{equation}
r(t) \sim \left(\frac{zt}{\sqrt{B}}\right)^{1/z}.
\end{equation}
Thus, at late times the particle follows a power-law trajectory with exponent $1/z$, which is characteristic of Lifshitz geometries and reduces to the familiar linear growth in the relativistic limit $z=1$. Substituting this asymptotic solution into the expression for the proper momentum gives
\begin{equation}
P_\rho(t) \sim m \left( \frac{zt}{\sqrt{B}\epsilon^z} \right)^{1-\frac{\theta}{dz}}.
\end{equation}
The power-law exponent of the proper momentum is therefore modified by the hyperscaling-violating exponent $\theta$, reflecting the additional scaling dimension introduced by this parameter. Consequently, the late-time growth of Krylov complexity takes the form
\begin{equation}\label{CKlateTOP}
C_K(t) \sim \frac{m\,\left(\sqrt{B}\,\epsilon^{z}\right)^{\theta/dz - 1}}{2z - \theta/d}\, (zt)^{2 - \theta/(dz)}.
\end{equation}
The above result coincides with the bottom-up result of Eq.~\eqref{CKlate} up to a constant factor, and shows that both $z$ and $\theta$ control the late-time growth exponent of Krylov complexity. For $\theta = 0$, the Lifshitz result is recovered, while for $\theta \neq 0$, the growth is either suppressed or enhanced depending on the sign of $\theta$. To summarise, in Table~\ref{tab:exponents} we report the complexity scaling exponents obtained from compactification on $S^{7-p}$ for various values of $p$. As a final remark, we note that for the expected values of the parameters, the condition $\theta = 2dz$ is never satisfied, and therefore there is no ambiguity in the result.

Interestingly, for $F-Dp$ bound states with $p = 5$, the critical exponents take special values that require separate treatment. This case is particularly interesting because the dynamical exponent vanishes and the particle exhibits qualitatively different behavior from the generic cases. From Eq.~\eqref{ztheta}, we see that at this point $z = 0$, $\theta = 8$, and $d = 6$. Hence, Eq.~\eqref{udot} becomes
\begin{equation}
\dot{r}=\frac{r}{\sqrt{B}}\sqrt{1-\left(\frac{r}{\epsilon}\right)^{8/3}}.
\end{equation}
For the velocity squared to be positive (real motion), we require $r \leq \epsilon$. However, the particle is released from rest at $r(0) = \epsilon$. Since $r$ cannot decrease — that would correspond to moving toward the boundary — the only consistent solution is $r(t) = \epsilon$ for all $t$. Thus, for $p=5$, the probe particle does not fall into the bulk and remains at the UV cutoff. A straightforward computation shows that the corresponding proper momentum vanishes, and hence the Krylov complexity growth rate vanishes identically, $\dot{C}_K(t) = 0$. As a result, the complexity remains zero for all times.

The case $F-D5$ bound states corresponds to a geometry where the effective potential prevents the probe particle from penetrating the bulk. This is a consequence of the fact that $\theta/d = 4/3 > 0$, which implies that the term $\left(r/\epsilon\right)^{2\theta/d}$ grows with $r$, making the square root in the velocity equation imaginary for $r > \epsilon$. Physically, this means that the geometry does not support radial infall of massive probes. Consequently, there is no complexity growth in the dual field theory according to the holographic prescription.

\section{Conclusions and discussions}\label{conclusion}

In this work, we have systematically investigated the holographic Krylov complexity in non-relativistic backgrounds characterized by Lifshitz scaling and hyperscaling violation. Following the proposal that the time derivative of Krylov complexity is identified with the proper radial momentum of an infalling massive probe, we have analyzed radial geodesic motion in both pure Lifshitz and hyperscaling-violating geometries.

For the pure Lifshitz case, we derived exact analytic expressions for the radial trajectory, the canonical momentum, and the proper momentum. The geodesic solution takes the simple form \( r(t) = ( \epsilon^{2z} + z^2 t^2 )^{1/(2z)} \), which yields a proper momentum growing linearly in time, \( P_\rho(t) = (m z / \epsilon^z) t \). Consequently, Krylov complexity scales quadratically, \( C_K(t) \sim t^2 \), for all values of the dynamical exponent \( z \). Thus, the non-relativistic scaling symmetry does not constrain the functional form of complexity growth; its only effect is an overall \( z \)-dependent factor in the rate. To further support this holographic finding, we have performed a complementary field-theory analysis of Krylov complexity in Lifshitz scalar and Dirac theories. Expanding the autocorrelation function for small times and solving the Schr\"odinger equation perturbatively, we find the same universal quadratic growth, \( K_{\mathcal{O}}(t) \sim t^2 \), at early times. While the quadratic form is universal, the coefficient is theory-dependent and differs between bosonic and fermionic systems, reflecting the distinct spectral properties of scalar and Dirac fields. This field-theoretic result is consistent with our holographic computation, confirming that the early-time quadratic growth is a robust feature of Krylov complexity in Lifshitz theories, arising from the fundamental structure of the Krylov construction.

For the hyperscaling-violating generalization, the presence of \( \theta \) introduces new scaling dimensions that modify both the geodesic motion and the complexity growth. While an explicit closed-form solution for \( r(t) \) is not available for generic parameters, we derived exact expressions for the radial velocity and momenta in terms of hypergeometric functions, and extracted the asymptotic behavior analytically. At late times, the proper momentum scales as \( P_\rho \sim t^{1 - \theta/(d z)} \), revealing that the hyperscaling-violating exponent controls the growth exponent directly. For \( \theta = 0 \), the Lifshitz result is recovered, while for \( \theta = d z \), the proper momentum approaches a constant, corresponding to logarithmic complexity growth. The relativistic AdS limit \( (z=1,\theta=0) \) reproduces the known result \( P_\rho \sim t \).

To strengthen the holographic interpretation and address the limitations of the bottom-up approach, we extended our analysis to explicit top-down constructions derived from string/M-theory. We first reviewed a type IIB supergravity solution exhibiting Lifshitz-like fixed points with dynamical exponent $z = 3/2$ for the $(x,y)$ directions and a different scaling for the $w$ direction, arising from a D3-D7 system with a linearly varying $\Theta$-angle. This background provides a controlled holographic realisation of anisotropic scale invariance with a known microscopic interpretation. We analyzed the geodesic motion of a massive probe in this geometry and obtained the corresponding Krylov complexity growth. Since the anisotropy parameter is encoded in the coefficient of the spatial direction, a purely radial geodesic with all spatial coordinates held constant effectively reduces the induced metric to that of asymptotically AdS. Consequently, the proper momentum and complexity growth are identical to the relativistic case. In particular, the complexity grows quadratically in time, consistent with the bottom-up Lifshitz result, which shows that the dynamical exponent alone does not affect the growth rate.

We then considered a family of intersecting F-Dp brane configurations, obtained via U-dualities from the D1-D5 system, which yield Lifshitz-like spacetimes with hyperscaling violation. The near-horizon limits of these solutions give rise to metrics characterised by critical exponents $z$ and $\theta$. After compactification on the internal sphere and transformation to the Einstein frame, we derived the reduced $(p+3)$-dimensional metric and analysed the radial geodesic motion of a massive probe. The canonical and proper momenta were computed explicitly, and the early-time and late-time asymptotic behaviors were extracted analytically. The early-time growth of Krylov complexity in the top-down models coincides with the bottom-up result up to a constant factor. This confirms that the quadratic early-time growth is a robust feature of holographic Krylov complexity, independent of the specific embedding. For the late-time regime, the results again match the bottom-up result up to an overall constant.

A comparison between the bottom-up and top-down analyses reveals that the time scaling exponents of Krylov complexity are identical in both approaches. The only differences lie in overall prefactors that depend on the specific embedding parameters (such as $Q_1$, $Q_2$ charges of the D-branes, and the sphere compactification). This suggests that the scaling behaviour of Krylov complexity is a universal feature of non-relativistic holographic backgrounds, insensitive to the details of the UV completion. The top-down constructions, however, provide additional benefits: they offer a concrete microscopic interpretation, allow for the study of extended probes and internal-space dynamics, and provide a controlled framework for future investigations of symmetry-resolved complexity and finite-temperature effects.

Beyond the specific results, our work provides a concrete realization of the holographic Krylov complexity dictionary in non-relativistic settings. It demonstrates that the momentum-Krylov correspondence extends beyond conformal backgrounds and remains applicable even in geometries with causal pathologies, provided one restricts to purely radial probes. This is particularly relevant given that Lifshitz spacetimes exhibit null curvature singularities, trapping of non-radial null geodesics, and causal wedge degeneracies that complicate holographic reconstruction. Our analysis shows that the proper momentum prescription, being local along the radial geodesic, circumvents these obstructions. It is instructive to compare our results with the CV and CA holographic complexity proposals in the same backgrounds. For the zero-temperature Lifshitz geometry, both CV and CA complexity exhibit quadratic growth in time, with the dynamical exponent $z$ entering only in the overall coefficient~\cite{Alishahiha:2018tep}. This matches our finding for Krylov complexity, suggesting that quadratic growth is a universal feature of quantum complexity in non-thermal states, independent of the specific holographic prescription. For hyperscaling-violating backgrounds, the late-time behavior of CV and CA complexity also shows $\theta$-dependent modifications~\cite{Alishahiha:2018tep}, analogous to our result $C_K \sim t^{2-\theta/(dz)}$. However, unlike CV and CA, the Krylov complexity prescription is intrinsically state-independent and free from ambiguities associated with reference states or boundary terms~\cite{Caputa:2024sux}, making it a more direct probe of operator growth dynamics in the dual field theory.

As mentioned earlier, Krylov complexity has also been studied directly in non-relativistic quantum field theories. In Refs.~\cite{Vasli:2023,Imani:2025etp}, various aspects of this measure were analyzed for free bosonic and fermionic field theories with Lifshitz scaling symmetry. In particular, the authors found that for a continuum massless Lifshitz scalar theory, Krylov complexity exhibits exponential growth in time, with the growth rate decreasing as the dynamical exponent $z$ is increased. Interestingly, the late-time slope remains independent of $z$. Similar behavior was observed in the fermionic case. It is worth noting that these field-theoretic results were obtained in free field theories at finite temperature, a regime where the holographic prescription appears to break down or require significant modification. In contrast, holographic duals typically describe strongly coupled field theories, and our computations are performed at zero temperature. This highlights the complementary nature of the two approaches: holographic computations in strongly coupled regimes at zero temperature yield power-law or logarithmic growth, while direct field-theory analyses in free theories at finite temperature exhibit exponential behavior. A complete understanding of Krylov complexity in Lifshitz theories would likely require bridging these different regimes of coupling and temperature.

There are several promising directions for future research. First, incorporating probes with internal structure or extended operators would provide a richer understanding of how non-relativistic symmetries affect symmetry-resolved and operator-dependent complexity. The top-down constructions reviewed here, with their explicit internal spaces, provide a natural setting for such extensions. Second, investigating the connection between Krylov complexity and other quantum information measures, such as entanglement entropy or circuit complexity, in Lifshitz holography would help establish a more complete picture of information dynamics in non-relativistic quantum field theories. Third, it would be interesting to explore whether the oscillatory behavior of Krylov complexity, recently identified as a universal signature of confinement, also appears in certain classes of Lifshitz or hyperscaling-violating geometries with an IR end-of-space. Finally, extending the top-down analysis to finite temperature and to probes with internal dynamics would provide a more complete holographic description of Krylov complexity in non-relativistic systems.

Overall, our results establish a systematic framework for computing holographic Krylov complexity in non-relativistic backgrounds and highlight the rich interplay between scaling symmetries, geodesic motion, and operator growth. The consistency between bottom-up and top-down approaches reinforces the robustness of the momentum-Krylov correspondence and its applicability beyond the conformal regime.

\subsection*{Acknowledgements}

We are very grateful to Carlos Nunez for correspondence, careful reading of the manuscript and his valuable comments.

\end{document}